\documentclass[12pt]{article}
\usepackage{graphicx}
\begin{document}

\title{MEASUREMENT OF THE PROTON ELECTROMAGNETIC FORM FACTORS AT BABAR}

\author{{\Large V.~P.~Druzhinin, on behalf of the BABAR Collaboration}
\vspace*{2mm}
\\
\it Budker Institute of Nuclear Physics SB RAS, Novosibirsk 630090,\\
\it Novosibirsk State University, Novosibirsk 630090, Russia
}
\date{} 
\maketitle

\begin{abstract}
The process $e^+e^-\to p\bar{p}$ has been studied in the $p\bar{p}$ mass
range from threshold to 6.5 GeV/$c^2$ using the initial-state-radiation technique
with both detected and undetected photon. The analysis is based on 
469 fb$^{-1}$ of 
integrated luminosity collected with the BABAR detector
at the PEP-II collider at $e^+e^-$ center-of-mass energies near 10.6 GeV.
\end{abstract}

\section{Introduction}
The energy dependence of the $e^+e^-\to p\bar{p}$ cross section is given by
\begin{equation}
\sigma_{p\bar{p}}(M_{p\bar{p}}) = \frac{4\pi\alpha^{2}\beta C}{3M_{p\bar{p}}^2}
\left [|G_M(M_{p\bar{p}})|^{2} + \frac{\tau}{2}|G_E(M_{p\bar{p}})|^{2}\right],
\label{eq1}
\end{equation}
where $M_{p\bar{p}}$ is the $p\bar{p}$ invariant mass,
$\tau=4m_p^2/M_{p\bar{p}}^2$, $\beta =\sqrt{1-\tau}$, 
$C=y/(1-e^{-y})$ is the Coulomb correction factor~\cite{Coulomb},
and $y = {\pi\alpha}(1+\beta^2)/\beta$. The cross section
depends on two form factors, electric $G_E$ and magnetic $G_M$.
From measurement of the cross section we determine 
the effective form factor
\begin{equation}
F_p(M_{p\bar{p}})^2=(|G_M(M_{p\bar{p}})|^{2} + 
\frac{\tau}{2}|G_E(M_{p\bar{p}})|^{2})/(1+\frac{\tau}{2}).
\end{equation}
It should be noted that such a definition was used in all previous 
measurements made with assumption that $G_E$ is equal to $G_M$. 

The $G_M$ and $G_E$ terms in the differential cross section have different 
angular dependencies, $1+\cos^2\theta$ and $\sin^2\theta$, respectively.
The ratio of the form factors can be extracted from the analysis of the
proton angular distribution. At threshold $G_E=G_M$, 
and the angular distribution is uniform.

The process $e^+e^-\to p\bar{p}$ is studied during 40 years
~\cite{ADONE73,DM1,DM2,FENICE,BES,CLEO,BABAR0,NU,BABAR1,BABAR2}.
However, all data before recent BABAR~\cite{BABAR0,BABAR1,BABAR2} and 
CLEO measurements~\cite{CLEO,NU}
had an accuracy of 20-30\%. The statistics was not sufficient to 
determine $G_E/G_M$ ratio from angular analysis. 

More precise results were obtained in the inverse reaction 
$p\bar{p}\to e^+e^-$~\cite{LEAR,E760,E835}. In PS170 experiment~\cite{LEAR} 
at LEAR the proton 
form factor was measured near threshold. A steep near-threshold mass-dependence
was observed. The $G_E$ to $G_M$ ratio was measured with about 30\% accuracy 
and was found to be compatible with unity. Above 3 GeV/$c^2$ measurements were
performed at Fermilab~\cite{E760,E835}. The strong decrease of the form 
factor was observed 
which agrees with the dependence $M_{p\bar{p}}^{-4}$ predicted by QCD for 
asymptotic proton form factor. Recently, very precise measurement of the 
form factor was performed on about 1.4 fb$^{-1}$ data collected by CLEO at 
3.77 and 4.17 GeV~\cite{NU}.
 
\section{Initial state radiation technique}
The initial-state-radiation (ISR) method is used at BABAR to measure
the $e^+e^-\to p\bar{p}$ cross section. In the ISR reaction
$e^+e^-\to p\bar{p}\gamma$ the photon is emitted by the initial electron or 
positron.
The mass spectrum of the $p\bar{p}$ pair is related to the cross section
of the nonradiative process $e^+e^-\to p\bar{p}$.

The ISR photons are emitted predominantly along beam axis.
There are two approaches in ISR measurements: tagged or large-angle (LA) ISR,
when the ISR photon is required to be detected,
and mainly small-angle (SA) untagged ISR. Only about
10\% of the ISR photons can be detected at BABAR calorimeter.
The produced ${p\bar{p}}$ system is boosted against the ISR photon.
Due to limited detector acceptance, the $M_{p\bar{p}}$ region below 
3 GeV/$c^2$ can be studied only with detected photon.
Above 3 GeV/$c^2$ statistics can be significantly increased with the use of
SA ISR. 

The advantage of the ISR method over conventional $e^+e^-$ and
$p\bar{p}$ experiments is that a wide mass region is studied in
a single experiment. 
The large-angle ISR has additional advantages. The first of them is
a low dependence of the detection efficiency on
the ${p\bar{p}}$ invariant mass. Measurement near and above threshold can 
be done with the same selection criteria.
The second is a low dependence of the detection efficiency on
hadron angular distributions (in the hadron rest frame). For protons this
significantly increases sensitivity for measurements of the $G_E/G_M$ ratio
and decreases model uncertainty in the cross section measurement.

Here we present BABAR results based on analysis of 469 fb$^{-1}$
data collected at $e^+e^-$ c.m. energy near 10.6 GeV.
Both LA ISR events~\cite{BABAR1} and SA ISR events~\cite{BABAR2}
have been used for analysis.

The selection of $e^+e^- \to p\bar{p}\gamma$ candidates requires
detection of two charged tracks of opposite charge originating from
the interaction region and identified as protons.
In LA ISR events a photon with the energy higher than 3 GeV
is additionally required. For each LA ISR candidate a kinematic fit with 
requirement of total energy and momentum conservation is performed.
The final selection is based on a condition on $\chi^2$ of the kinematic fit.
For SA ISR candidate the selection is based on two parameters: the transverse
momentum of the $p\bar{p}$ pair, and the invariant mass of the 
system recoiling against $p\bar{p}$. Both parameters 
should be close to zero.

The dominant source of background in the LA case arises from  
$e^+e^-\to p\bar{p}\pi^0$ events with an undetected low-energy photon, or with
merged photons from the $\pi^0$ decay. This background is estimated  
using a control sample of $e^+e^-\to p\bar{p}\pi^0$ events. Its level is found
to change from 5\% near threshold to 50\% at 4 GeV/$c^2$. All observed 
$p\bar{p}\gamma$ candidates with the mass greater than 4.5 GeV/$c^2$ are 
consistent with  $p\bar{p}\pi^0$ background. The contribution of other 
background processes is estimated to be about 1\% of selected data events.
The dominant background sources in the SA ISR case are
the ISR process $e^+e^- \to p\bar{p}\pi^0\gamma$ and 
two-photon $p\bar{p}$ production. The background level is estimated
to be about 5\% and subtracted.

\section{Results}
\begin{figure}
\centering{
\includegraphics[width=8.75cm]{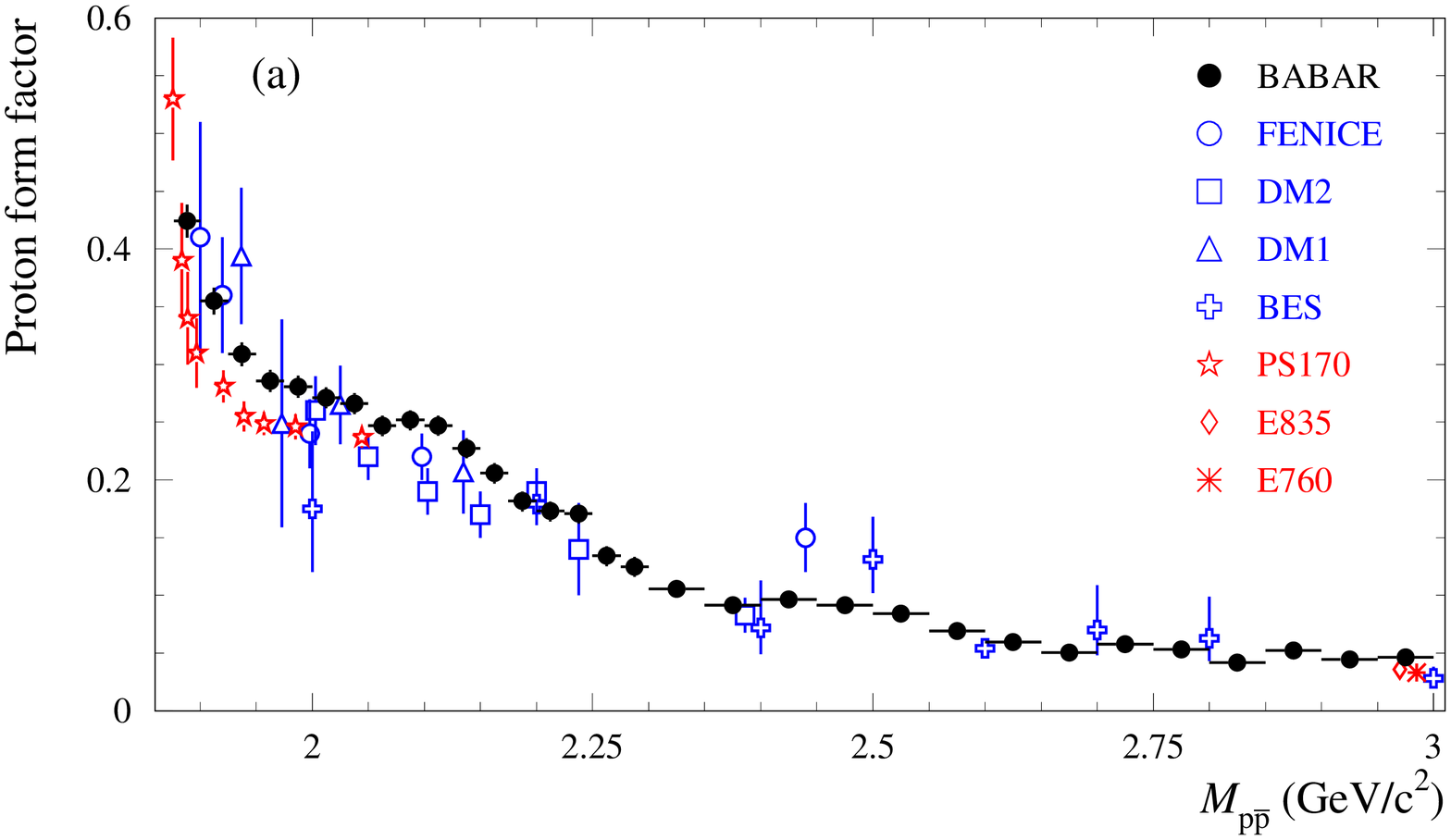}
\includegraphics[width=7cm]{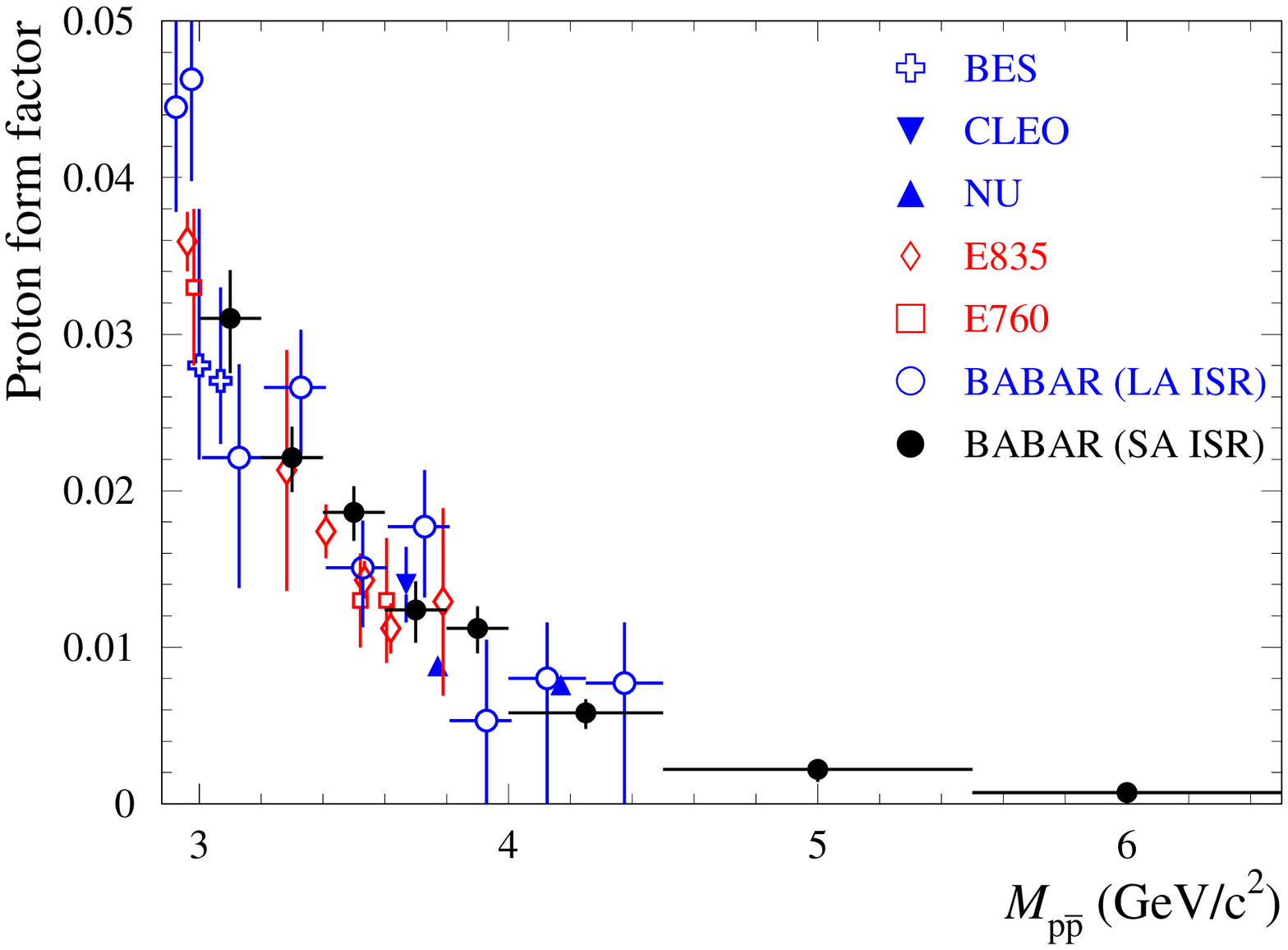}}
\vspace*{-10pt}
\caption{
The mass dependence
of the effective proton form factor in two different mass region measured 
by BABAR in comparison with results of previous experiments.
\label{fig1}}
\end{figure}
From the measured $p\bar{p}$ mass spectrum we obtain the 
$e^+e^-\to p\bar{p}$ cross section and the proton effective form factor.
In the mass region under study, the cross section changes by 6 orders 
of magnitude, from about 1 nb at the $p\bar{p}$ threshold to about 1 fb at 
6 GeV/$c^2$. The measured effective form factor is shown in Fig.~\ref{fig1}
in comparison with existing $e^+e^-$ and $p\bar{p}$ data. 
Our data are in reasonable agreement with previous measurements everywhere
except near-threshold region, where the BABAR results are systematically 
larger than the PS170 data~\cite{LEAR}.

\begin{figure}
\centering{\includegraphics[width=6cm]{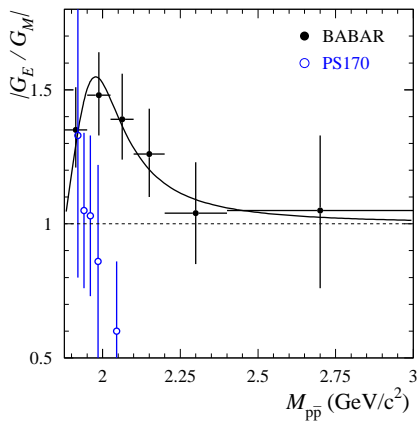}}
\vspace*{-10pt}
\caption{The $|G_E/G_M|$ mass dependence measured by BABAR
in comparison with PS170 data. 
\label{fig1a}}
\end{figure}
The measured mass dependence of the ratio $|G_E/G_M|$ is shown in 
Fig.~\ref{fig1a}. To measure the ratio the distribution of $\cos{\theta_p}$ is
analyzed, where $\theta_p$ is the angle between the proton momentum in
the $p\bar{p}$ rest frame and the momentum of the $p\bar{p}$ system
in the $e^+e^-$ c.m. frame. The measured ratio is higher than unity at 
masses below 2.2 GeV/$c^2$. Our results disagree with the previous PS170 
measurement~\cite{LEAR}.

We have also searched for an asymmetry in the proton angular distribution.
An asymmetry is absent in the lowest order (one-photon $p\bar{p}$ production).
It arises from higher-order contributions (soft extra ISR and FSR interference,
two-photon exchange). The integral asymmetry for events with
$p\bar{p}$ mass below 3 GeV/$c^2$ is found to be consistent with zero:
\begin{equation}
A_{\cos{\theta_p}}=\frac{\sigma(\cos{\theta_p}>0)-\sigma(\cos{\theta_p}<0)}
{\sigma(\cos{\theta_p}>0)+\sigma(\cos{\theta_p}<0)}=-0.025\pm0.014\pm0.003.
\end{equation}

\begin{figure}
\begin{center}
\includegraphics[width=5cm]{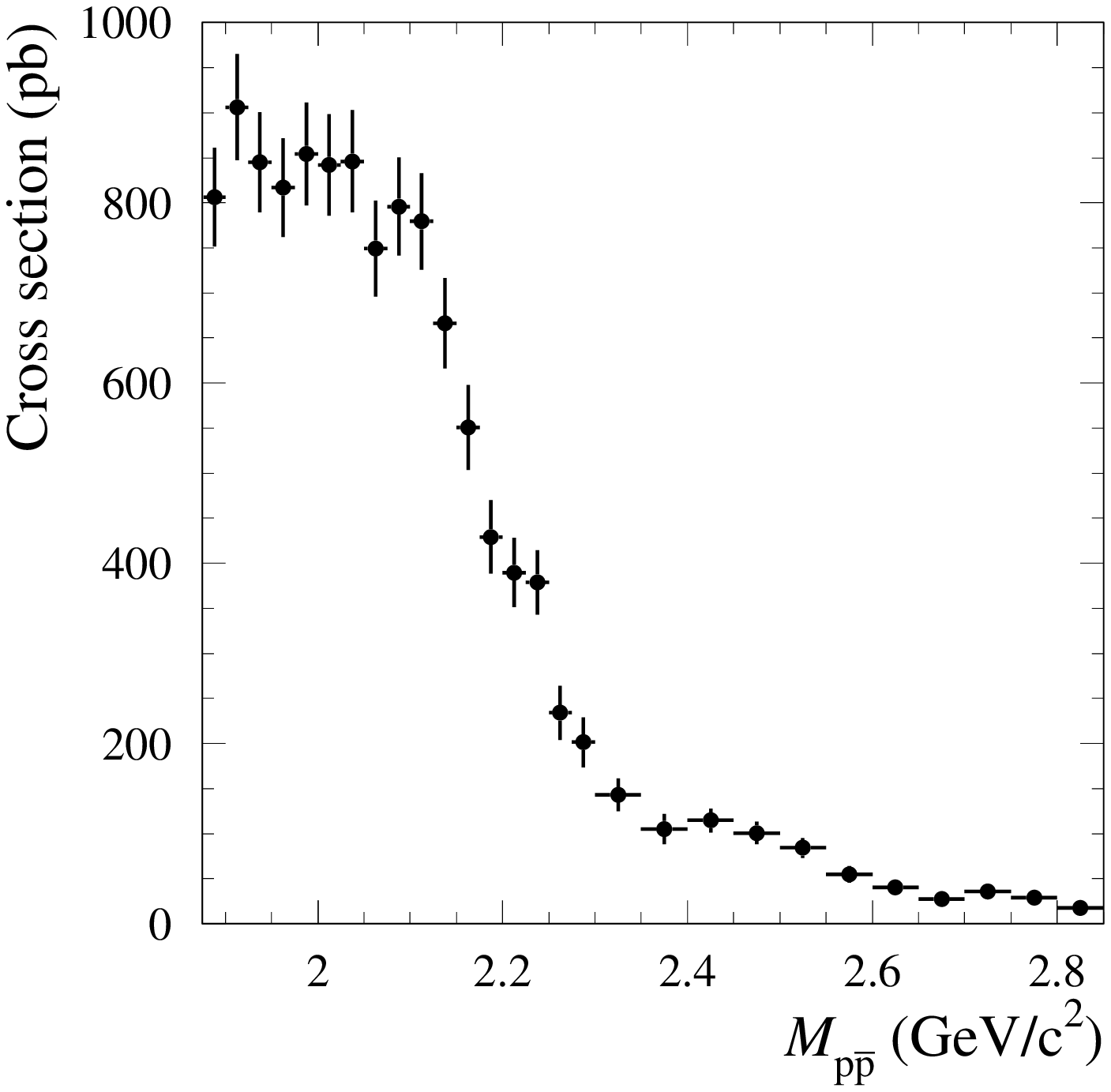}
\includegraphics[width=5cm]{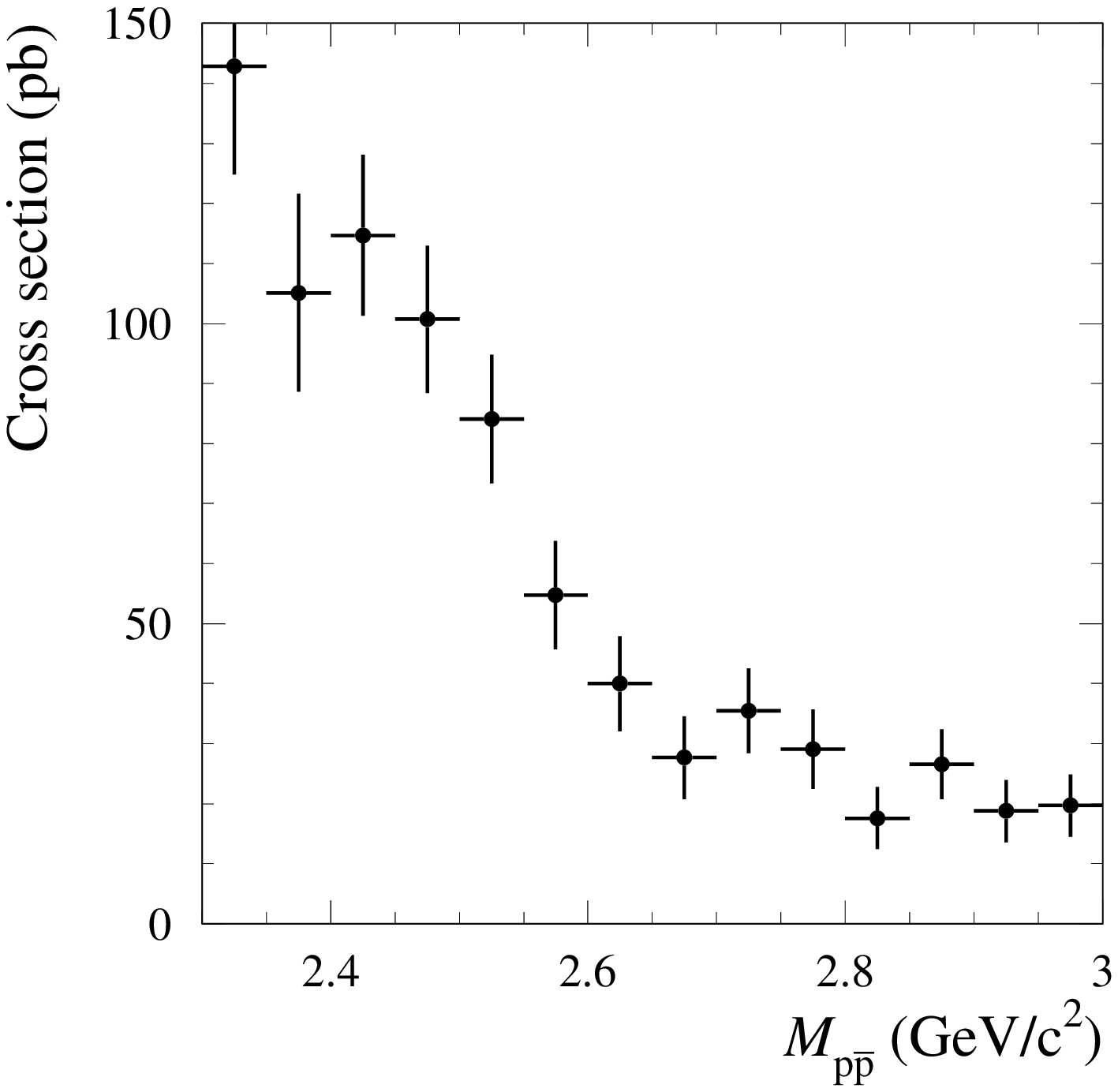}
\includegraphics[width=5cm]{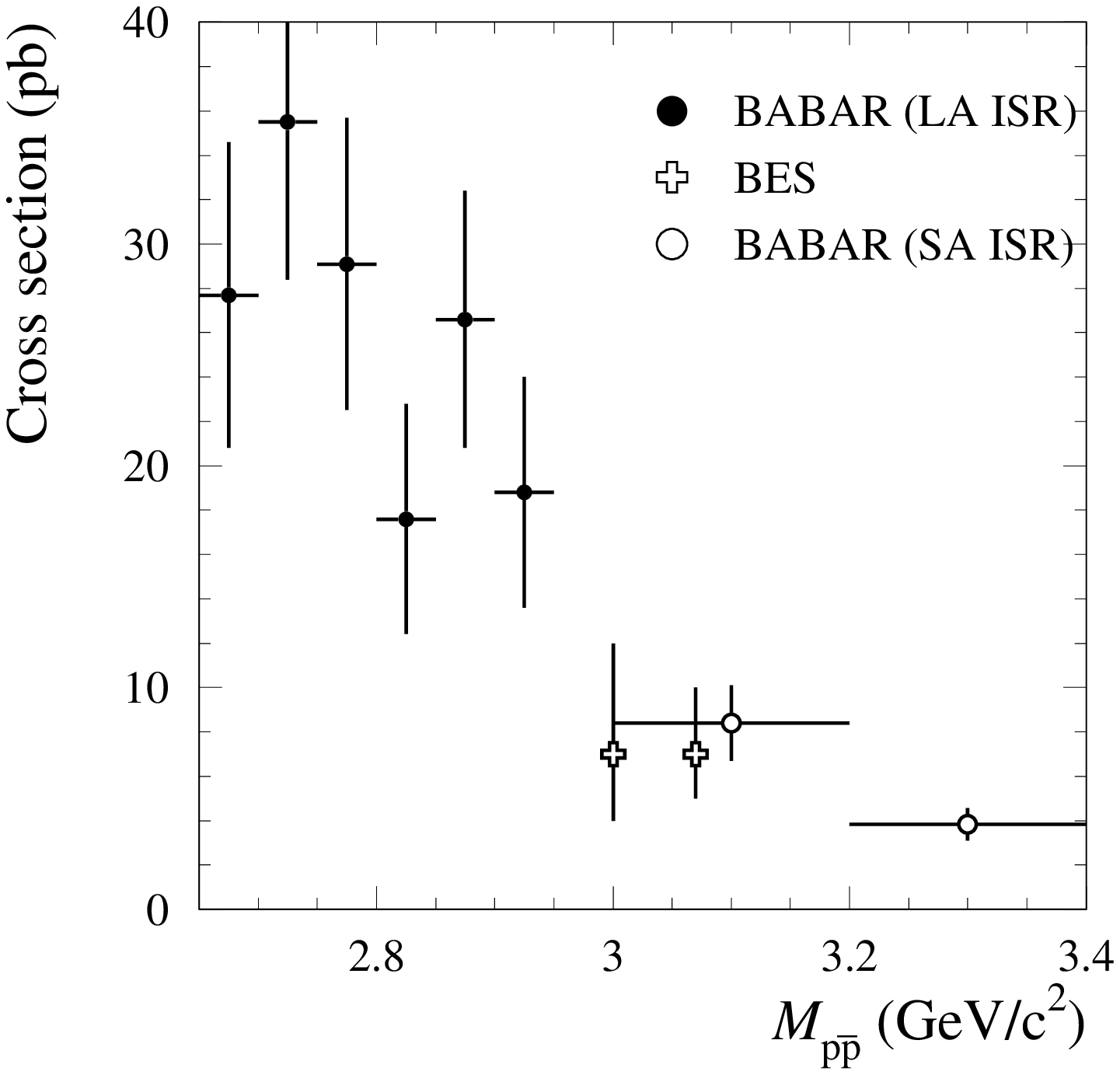}
\end{center}
\vspace*{-10pt}
\caption{The $e^+e^-\to p\bar{p}$ cross section in the mass regions
near 2.2 (left), 2.5 (middle), and 3 (right) GeV/$c^2$.
\label{fig2}}
\end{figure}
The measured form factor has a complex mass dependence.
The growth of the form factor near threshold
as well as the deviation of the ratio $|G_E/G_M|$ from unity
may be due to final-state interaction between the proton and
antiproton~\cite{dmitriev}. At higher energies the form factor
and cross section display a steplike mass dependence with
three steps near 2.2, 2.5, and 3 GeV/$c^2$. Such a dependence
is not described by existing models for the form factors
(see, for example, Refs.~\cite{ffmod1,ffmod2,ffmod3,ffmod4}).
The $e^+e^-\to p\bar{p}$ cross section in the mass regions of the steps
is shown in Fig.~\ref{fig2}.

\begin{figure}
\begin{center}
\includegraphics[width=6.2cm]{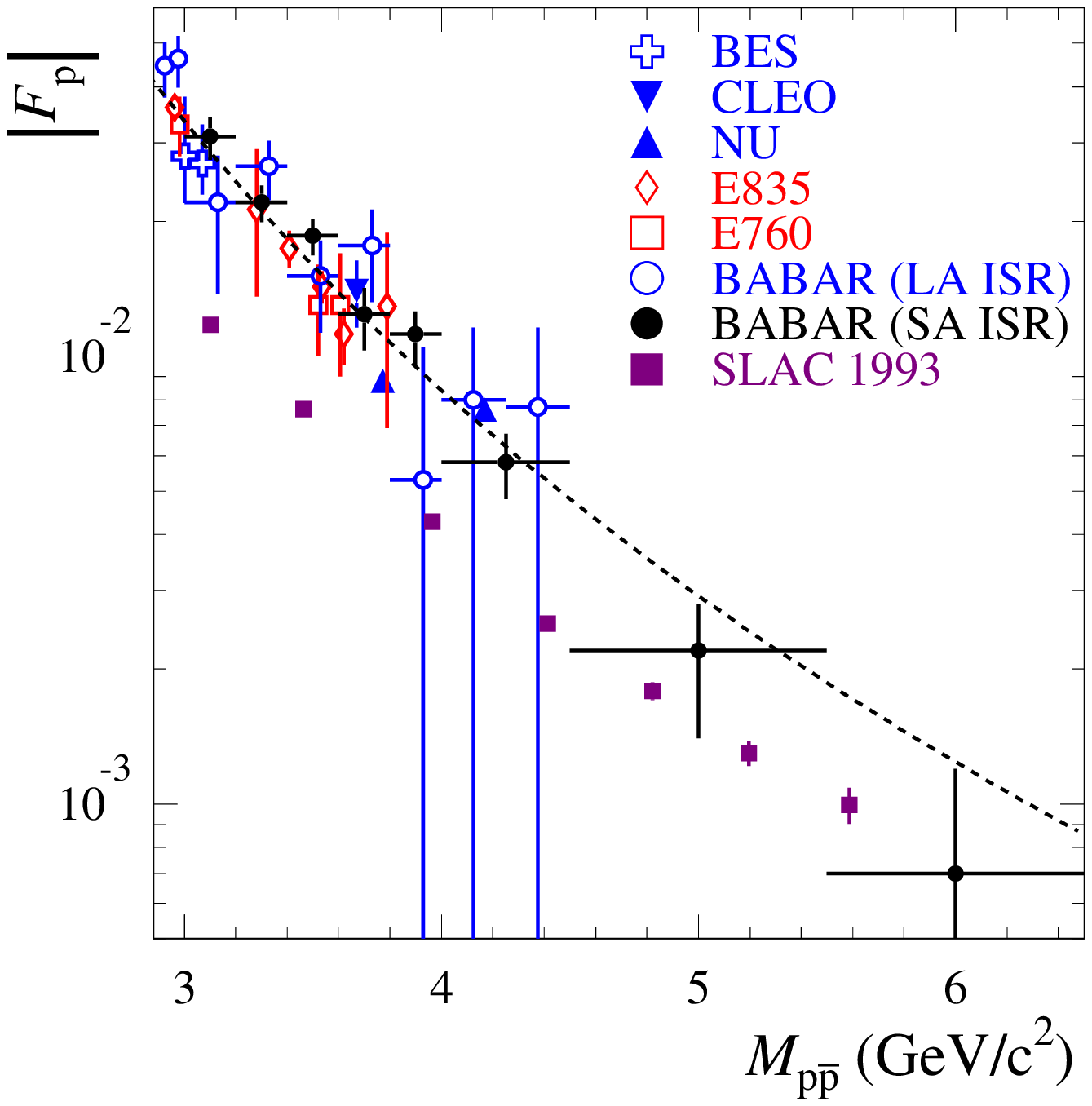}
\includegraphics[width=8.267cm]{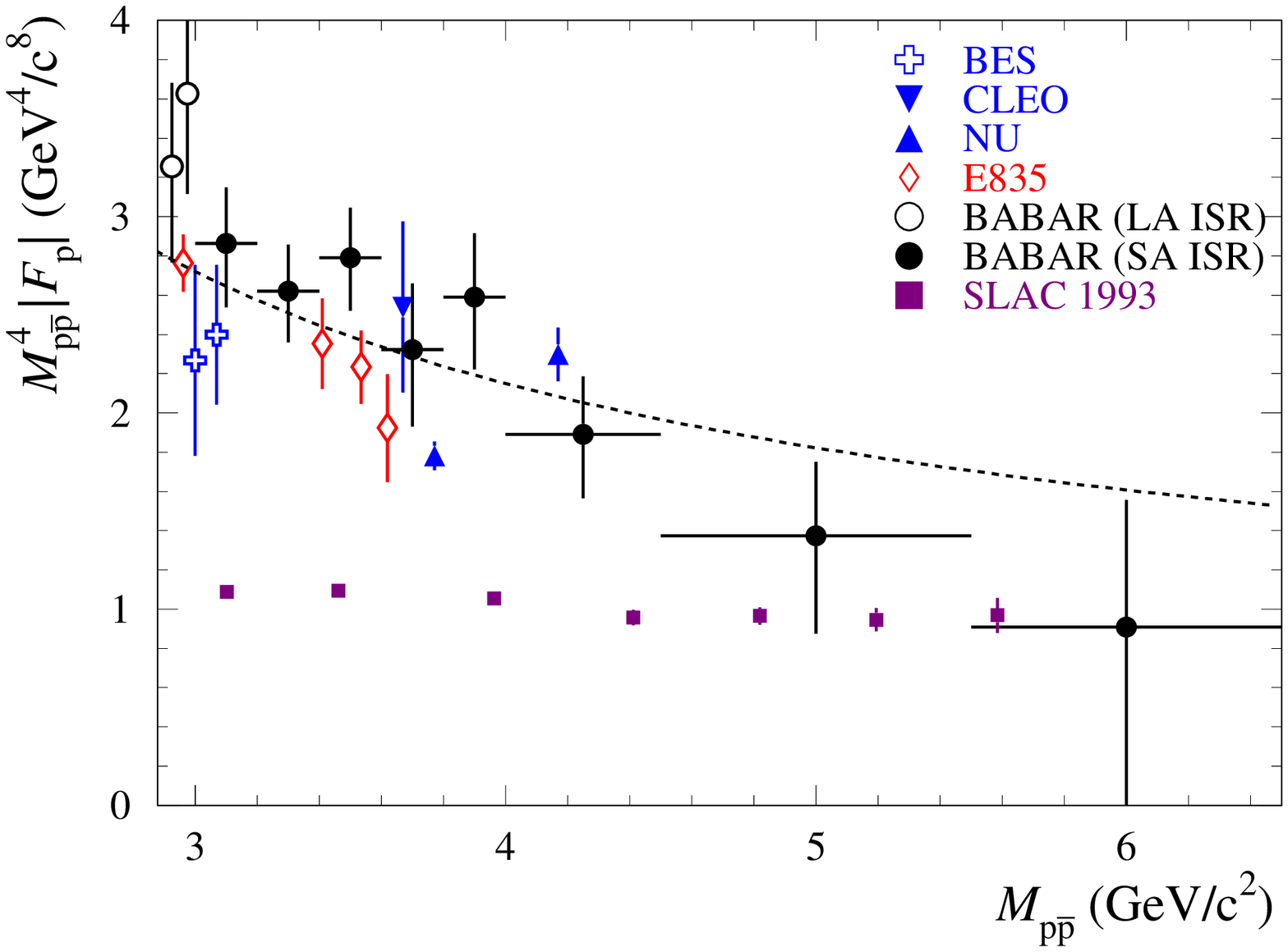}
\end{center}
\vspace*{-10pt}
\caption{Left plot: the proton effective form factor above 3 GeV/$c^2$ in
log scale. Right plot: the scaled (multiplied by $M_{p\bar{p}}^4$)
form factor. In the right plot the data points with largest errors were
excluded. The points denoted by ''SLAC 1993'' represent data on the space-like
magnetic form factor. The curve is the result of the QCD-motivated fit.
\label{fig3}}
\end{figure}
Figure~\ref{fig3} depicts the existing form-factor data above
3 GeV/$c^2$ in log scale. To compensate the main mass dependence of the form
factor ($1/m^4$) we also show the scaled (multiplied by $M_{p\bar{p}}^4$) 
form factor.  The dashed curve in Fig.~\ref{fig3}
corresponds to a fit of the asymptotic QCD dependence of
the proton form factor, 
$m^4F_p\sim \alpha_s^2(m)$~\cite{QCD},
to the form factor data.
All the data above 3 GeV/$c^2$ except the two points marked ``NU''~\cite{NU} are 
well described by this function. Adding the
``NU'' points changes the fit $\chi^2/\nu$ from
17/24 to 54/26.
Our data shows that the form factor decreases in agreement
with the asymptotic QCD prediction.  The decrease may be
even faster above 4.5 GeV/$c^2$.
The local deviations of the ``NU'' points from the global fit may be
result of the $\psi(3770)$ and $\psi(4160)$ resonance contributions.

The points marked ``SLAC 1993'' represent data on the space-like magnetic
form factor measured in $ep$ scattering~\cite{SLAC}.
The asymptotic values of the space- and time-like form factors are
expected to be the same.
In the mass region from 3 to 4.5 GeV/$c^2$ the time-like form factor is about 
two-three times larger than the space-like one.
The new BABAR data at high masses give an indication that the difference
between the time- and space-like form factors decreases with mass increase.
\section{Summary}
The $e^+e^-\to p\bar{p}$ cross section and the proton effective form factor
have been measured from threshold up to 6.5 GeV/$c^2$ using the full BABAR data
sample.

The form factor has complex mass dependence. There are a near-threshold steep
falloff and a step-like behavior at higher masses.
At masses above 3 GeV/$c^2$ the observed decrease of the form factor agrees 
with the asymptotic dependence predicted by QCD or is even faster.

The $|G_E/G_M|$ ratio has been measured from threshold to 3 GeV/$c^2$.
A large deviation of this ratio from unity is observed below 2.2 GeV/$c^2$.
The asymmetry in the proton angular distribution has also been measured.

\end{document}